\documentclass[a4paper]{emulateapj} 
\usepackage{graphicx}

\def\gr{$\gamma$-ray} 
\def\ls {LS~5039}
\def\be{O6.5V }
 
\shorttitle{TeV emission from \gr -loud binaies} 
\shortauthors{Neronov \& Chernyakova} 
 
\begin{document} 
\title{A rotating hollow cone anisotropy of TeV emission from binary
systems} 
\author{A.Neronov} 
\affil{{\it INTEGRAL} Science Data Center,
16 ch. d'Ecogia, CH-1290, Versoix, Switzerland}
\email{andrii.neronov@obs.unige.ch}
\and
\author{M.Chernyakova}
\affil{Dublin Institute for Advanced Studies,31 Fitzwilliam Place,
Dublin 2, Ireland } 
\email{masha@cp.dias.ie}                   
\begin{abstract}  
  We show that TeV \gr\ emission produced via interactions of
  high-energy particles with anisotropic radiation field of a massive
  star in binary systems should have a characteristic rotating hollow
  cone anisotropy pattern. The hollow cone, whose axis is directed
  away from the massive star, rotates with the period equal to the
  orbital period of the system. We note that the two maxima pattern of
  the TeV energy band lightcurve of the \gr\ loud binary LS 5039 can
  be interpreted in terms of this rotating hollow cone model.
  Adopting such an interpretation, we are able to constrain the
  geometry of the system -- either the inclination angle of the binary
  orbit, or the elevation of the \gr\ emission region above the
  orbital plane.
\end{abstract}

\maketitle

\textit{Introduction.}
\gr-loud binary systems are a newly identified class of sources in
which either accretion onto the compact object (a neutron star, or a
black hole), or interaction of an outflow from the compact object with
the wind and radiation from a massive companion star leads to the
production of very-high energy (VHE) \gr\ emission. Three such
systems, PSR B1259-63, LS 5039 and LSI +61 303, have been firmly detected
as persistent or regularly variable TeV \gr\ emitters
\citep{aharonian05,aharonian06,albert06}.
The VHE \gr\ emission from the \gr -loud binaries is variable on the
orbital period (or shorter) time scale. This implies that the emission
region is located close to the binary system, in a highly
inhomogeneous and anisotropic particle and photon background produced
by massive companion star.

In what follows we show that if the \gr\ emission from such a region
is produced in interactions of isotropically distributed VHE particles
with photons from the massive star, it should have a characteristic
``rotating hollow cone'' anisotropy, i.e. most of the photons are
emitted at a certain angle $\zeta_0$ with respect to a symmetry axis directed 
radially away from the massive star. Orbital motion of the emission region
around the massive star leads to the rotation of the emission cone.
Rotation of the hollow cone on the orbital time scale leads to
the appearance of 0, 1, or 2 maxima in the orbit-folded lightcurve in
the VHE band, occurring at the phases when the line of sight is
inclined at an angle $\zeta_0$ with respect to the cone axis, i.e. at
the moments of passage of the of the cone through the line of sight
(similarly to the hollow cone models of period-folded lightcurves of
pulsars, see e.g. \citep{lyne05}).

The orbital modulation of the \gr\ flux, related to the passage of the
hollow cone through the line of sight could be most clearly detected
if there are no additional sources of the modulation, related e.g. to
the ellipticity of the binary orbit, absence of spherical symmetry of
the wind/radiation from the companion star etc. Among the three
\gr-loud binary systems mentioned above, the system LS 5039 is
characterized by the lowest ellipticity of the orbit. In this system
the compact object orbits a \be\ star which emits isotropic stellar
wind (contrary to the other two systems in which the massive star is
of the Be type). 

The influence of the anisotropy of the photon field of the massive star on the
properties of the \gr\ emission in binaries in general, and in LS 5039 in
particular, was first studied by \citet{khangulyan05,khangulyan07}. Here we
calculate the angular brightness profile of the hollow cone in LS 5039, and
find that the observation of the two maxima of the orbit-folded lightcurve
constrains the inclination of the binary orbit to be $i>40^\circ$, if the
emission is produced in the vicinity of the compact object.  This result can be
stated also in an opposite way: if the inclination of the binary orbit is
$i<40^\circ$, the two maxima structure of the orbit-folded lightcurve can be
explained only if the VHE \gr\ emission region is displaced from the position
of the compact object. This can be the case if the emission is produced in a
jet. In this latter case, we show that the existence of the two-maxima of the
lightcurve constrains the elevation of the emission point above the orbital
plane.

\textit{Anisotropy of VHE $\gamma$-ray 
emission in a central photon field.}
Consider the \gr\ emission produced by interactions of VHE
particles $X$ (e.g. protons or electrons) with the soft photon field
in the vicinity of a massive star. Assume for simplicity that
the size of the emission region is much less than the
distance from the region to the center of the star and that the VHE
particles in the region have isotropic velocity distribution.
In spite of the isotropy of the VHE particle distribution, the \gr\
emission will be anisotropic.  The anisotropy arises because of the
Doppler effect which leads to the decrease (increase) of the rate of
interaction of the VHE particles co-moving with (moving oppositely to)
the soft photon field of the massive star.

The interaction rate of particles $X$ with momenta ${\bf P}_X$ with
soft photons with momenta ${\bf p}_*$ is given by 
\citep{landau80}
\begin{equation}
\label{eq:R}
{\cal R}=\int n_{X}n_*\sigma
\frac{{\bf P}_X\cdot {\bf p}_*}{E_X \epsilon_*}
dE_{X} d\epsilon_* d\Omega\ ,
\end{equation}
where $\sigma$ is the interaction cross-section, $n_{X}({\bf P}_X)$ is
the particle distribution, $n_*({\bf p}_*)$ is the soft photon
distribution and ${\bf P}_X\cdot {\bf p}_*$ is the scalar product of
the 4-momenta of the interacting particles,
\begin{equation}
{\bf P}_X\cdot {\bf p}_*\simeq E_x\epsilon_*(1-\cos\zeta)\ ,
\end{equation}
($\zeta$ is the angle between the particle velocities) which contains
the Doppler factor, $(1-\beta \cos\zeta)$ (we assume that particle
velocity is $\beta\simeq 1$).

If the soft photon field in the emission region would be isotropic,
integration of the interaction rate of Eq. (\ref{eq:R}) over the soft
photon angular distribution would average out of the angular factor
$(1-\cos\zeta)$, so that the \gr\ emission intensity, which is
proportional to the interaction rate, would be isotropic. However,
since all the soft photons crossing the emission region move in the
same direction (away from the massive star), the interaction rate
depends on the angle $\zeta$ between the direction of emission and the
direction "from the massive star". Particles $X$ comoving with the
soft photon field (moving at the angles $\zeta\rightarrow 0^\circ$)
interact more rarely than the particles moving opposite to the photon
field (at $\zeta\rightarrow 180^\circ$).

The reduction of the interaction rate has two-fold consequences. On
one hand, it leads to the reduction of the power of the \gr\ emission
by the particles moving in the direction away from the star.  On the
other hand, the reduction of the interaction rate of the soft photons
with the emitted VHE \gr s facilitates the escape of \gr s moving in
the direction away from the star. A competition between the decrease
of the \gr\ production rate, ${\cal R}_{\rm prod}(\zeta)$, and the
increase of the \gr\ ``survival probability'' (i.e. of the
$\exp\left(-\tau_{\rm abs}(\zeta)\right)$, where $\tau_{\rm
  abs}(\zeta)$ is the optical depth with respect to the pair
production) leads to the appearance of a maximum of the \gr\ flux
\begin{equation}
\label{F}
F_\gamma(\zeta)\sim {\cal R}_{\rm prod}(\zeta)e^{-\tau_{\rm abs}(\zeta)}
\end{equation}
at an angle $0<\zeta_0<\pi$,
i.e. to the appearance of a hollow cone anisotropy pattern.

At large distances from the massive star one can approximate the
angular distribution of the soft photons by that of a point source.
Under this simplifying assumption, one finds that the integration over
the angular distribution of the UV photons is easily performed and the
resulting expression for the rate of production of \gr s takes the
form
\begin{equation}
\label{rate}
{\cal R}_{\rm prod}(\zeta)\simeq n_Xn_*\sigma_{\rm prod}(1-\cos\zeta)\ ,
\end{equation}
where $\sigma_{\rm prod}$ is the cross-section of production of \gr s in
interaction of the particles $X$ with the soft photon field. 

An estimate of the optical depth for the \gr s, escaping from the
production region, can be obtained by multiplying the absorption rate
per \gr\ on the size of the absorbing region. The absorption rate
 is given by the expression (\ref{eq:R}) in which
the particle $X$ is a \gr\ and the cross-section $\sigma$ is the pair
production cross section, $\sigma_{\rm abs}$.  Estimating the size of
the absorption region to be of the order of the distance $D$ of the
emission point from the massive star, one finds
\begin{equation}
\label{tauabs}
\tau_{\rm abs}\simeq {\cal R}_{\rm abs}D/n_\gamma
\sim  n_*\sigma_{\rm abs}D(1-\cos\zeta)\ .
\end{equation}

Substituting Eqs. (\ref{rate}),(\ref{tauabs}) into Eq. (\ref{F})
one finds
\begin{equation}
\label{Fzeta}
F_{\gamma}(\zeta)\sim
\sigma_{\rm prod}(1-\cos\zeta)e^{
-n_*D\sigma_{\rm abs}(1-\cos\zeta)}\ .
\end{equation}
Two effects affect the anisotropy pattern of the \gr\ emission. First, the
explicit dependence of $F_\gamma$ on $\cos\zeta$ is introduced in Eq.
(\ref{Fzeta}) by the Doppler effect. 
An additional implicit dependence on $\zeta$ is introduced  
through the energy dependence of the interaction cross-sections $\sigma_{\rm
prod},\sigma_{\rm abs}$. Indeed, in
general 
$\sigma=\sigma(E_{\rm CM})$, where the $E_{\rm CM}$ if the center-of-mass energy,
which, in the case $E_{\rm CM}\gg m_X$ depends on $\zeta$ as $E_{\rm CM}\sim
\sqrt{1-\cos\zeta}$. 

\begin{figure}
\plotone{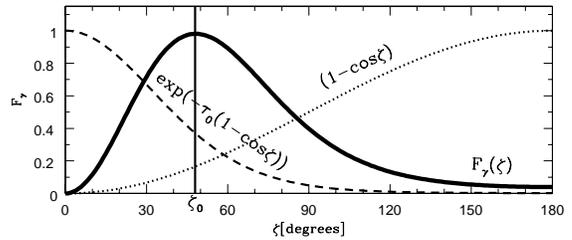}
\caption{Thick solid curve: angular brightness profile, 
Eq. (\ref{Fzeta}), in the approximation of constant
cross-sections. Thin dotted and dashed curves show the 
\gr\ production rate and absorption coefficients as functions of the angle
$\zeta$.}
\label{fig:Fzeta}
\end{figure}

The anisotropy pattern resulting from the Doppler effect can be found
if one ignores the energy dependence of $\sigma_{\rm prod},
\sigma_{\rm abs}$. This is done in Fig. \ref{fig:Fzeta} (the constant
$\tau_0=n_*D\sigma_{\rm abs}$ is taken to be $\tau_0=3$). From this
figure one can see that most of the \gr\ flux is emitted along a
"thick hollow cone" with the opening angle $\zeta_0$ and the thickness
comparable to the opening angle, $\Delta\zeta\sim \zeta_0$. The energy
dependence of the interaction cross-sections leads to the
energy-dependent angular brightness profile of the thick hollow
cone. The cone becomes wider at higher energies, where the absorption
is less efficient, see Fig. \ref{fig:Fzeta_num} below. The non-zero
brightness at $\zeta=0$ is due to the finite radius of the star.

\begin{figure}
\begin{center}
\plotone{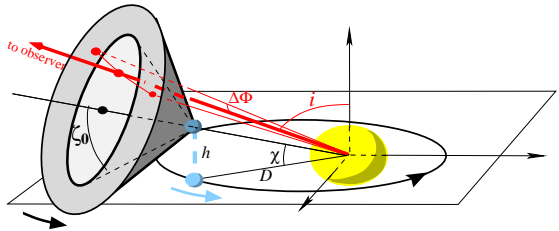}
\caption{Model of the anisotropy of the \gr\ emission. 
Maxima of the lightcurve are expected
  during the passage of the line of sight through the hollow cone,
  whose axis is directed along the line connecting the emission region. }
\label{fig:conus}
\end{center}
\end{figure}

\textit{ Geometrical model of variability of $\gamma$-ray emission.}
If the location of the emission region is determined by the position
of the compact object (e.g. the \gr\ emission is produced in the
vicinity of the compact object, or in the jet emitted by the compact
object), the orientation of the hollow cone changes when the compact
object moves around the star, as it is shown in
Fig. \ref{fig:conus} (where the emission region is supposed to be
situated at an elevation $h$ above the compact object, so that the
hollow cone axis is inclined at an angle $\chi=\mbox{atan}(h/D)$,
where $D$ is the binary separation distance).

Changes of the orientation of the hollow cone with respect to the line of sight
should lead to the orbital modulation of the observed \gr\ flux. 
Depending on the relation between the inclination angle $i$ of
the binary orbit plane, and the opening angle of the hollow cone,
$\zeta_0$, two characteristic patterns of the orbital modulation are possible. 
If the inclination of the orbit is
$i>\pi/2-\zeta_0-\chi$, the direction
of the line of sight passes through the "wall" of the cone two times per
orbit. This should lead to the appearance of two maxima in the orbit-folded
lightcurve. If the binary orbit is circular, the maxima of the
lightcurve occur when the true anomaly of the orbit (angular orbital phase
$0<\Phi<360^\circ$, counted from the focal point of the orbit, $\Phi=0^\circ$ at
the periastron) takes the values
\begin{equation}
\label{phi12}
\Phi_{1,2}=\Phi_{\rm inf}\pm \Delta\Phi
\end{equation}
where $\Phi_{\rm inf}$ is the anomaly of the inferior conjunction and
\begin{equation}
\label{DP}
\Delta\Phi=\mbox{acos}\left[1-(\sin(i+\chi)-\cos\zeta_0)/(\cos\chi\sin i)\right]
\end{equation} 
If the binary orbit is elliptical, an additional orbital modulation of the
\gr\  flux can occur because of the variation of the distance of the emission
region from the massive star (which leads to the modulation of the \gr\
production/absorption rates) with the
orbital phase.  As the inclination of the orbit decreases to 
$i\le\pi/2-\zeta_0-\chi$, the two maxima at $\Phi_1$ and $\Phi_2$ merge at the
phase $\Phi_{\rm inf}$.

Since the opening angle of the hollow cone, $\zeta_0$ depends on the
\gr\ energy, the condition
$i>\pi/2-\zeta_0(E_\gamma)-\chi$ can be satisfied only in a certain
interval of energies, so that a merger of the two maxima
$\Phi_{1,2}=\Phi_{\rm inf}\pm\Delta\Phi(E_\gamma)$ at the phase of
inferior conjunction can be observable at a particular energy $E_0$ at
which $\zeta_0(E_0)=\pi/2-i-\chi$.

\textit{ The case of LS 5039.}
\ls\ is one of the several X-ray binaries detected as sources of the
VHE \gr\ emission \citep{aharonian06}. In this binary system the
compact object rotates with the period $P=3.9078\pm0.0015$~d around a
massive \be star. The orbit is eccentric ($e=0.35$). The inclination
of the orbit is $20^\circ<i<60^\circ$ \citep{casares05}.  

\begin{figure} 
\plotone{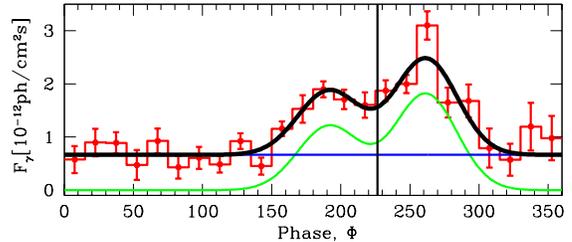} 
\caption{A
phenomenological model fit to the lightcurve of LS 5039. The model
consists of a constant plus two gaussians of equal width at equal
distance from the inferior conjunction (shown by a vertical line).}
\label{fig:lc_Phi} 
\end{figure}

The orbit-folded lightcurve of the source at the energies $E>1$~TeV
\citep{aharonian06} has two maxima at the orbital phases $\phi_1\simeq
0.55,\phi_2\simeq 0.85$. The two maxima are apparently not symmetric:
the first maximum around $\phi_1$ spans a broader range of the orbital
phase, while the second maximum is more sharp. The second maximum
happens closer to the phase of the inferior conjunction, $\phi_{\rm
  inf}\simeq 0.716$, than the first one.  Replotting the orbit-folded
TeV \gr\ lightcurve as a function of the true anomaly, $\Phi$, (to
produce Fig. \ref{fig:lc_Phi}, we have calculated the true anomalies
of each data point of the top panel of Fig. 5 of \citet{aharonian06}
and rebinned the data into bins of the width $d\Phi=15^\circ$) rather
than as a function of $\phi$ , a symmetry in the positions of the two
maxima can be found.

Namely, the lightcurve can be satisfactory fitted with a
phenomenological model which is a sum of a constant (weakly modulated
emission which can be produced e.g. at larger distances) plus two
gaussians, whose centers are equally spaced from the phase of the
inferior conjunction, $\Phi_{\rm inf}\simeq 224^\circ$ (see
Fig. \ref{fig:lc_Phi}).  The positions of the centers of the
gaussians, found by the fit, are $\Phi_{1,2}\simeq \Phi_{\rm inf}\pm
35^\circ$, while the widths of the gaussians are nearly equal,
$\delta\Phi_{1,2}\simeq 22.5^\circ$ (if the phase of the inferior
conjunction is left free, while fitting, the fit finds the phase
$\Phi_{\rm inf}\simeq 226.5^\circ$, consistent with the value
$\Phi_{\rm inf}\simeq 224.2^\circ\pm 3.3^\circ$ found by
\citet{casares05}). The gaussian centered at the phase $\Phi_2$ is
found to have $\simeq 1.5$ times higher normalization than the
gaussian centered at $\Phi_1$.

\begin{figure} 
\plotone{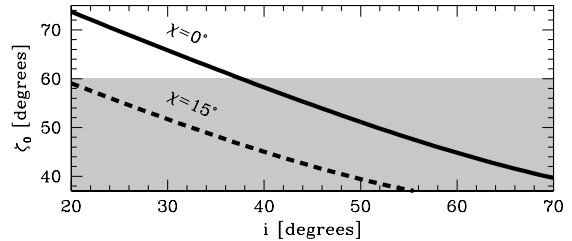}
\caption{Dependence of the opening angle of the hollow cone on the inclination
of the binary orbit in LS 5039. Grey shaded region indicates the range of
$\zeta_0$ found from numerical calculations.}
\label{fig:zeta_i}
\end{figure}

The observed symmetry of the positions of the maxima of the orbit folded
lightcurve can be interpreted, in a straightforward way, in terms of the
"rotating hollow cone" model, discussed in the previous sections. In particular 
the phase shift of Eq. (\ref{phi12}) is $\Delta\Phi\simeq 35^\circ$. 
From Eq. (\ref{DP}) (which can be used, for a low-eccentricity orbit, as a
first approximation) one can find  a relation between $i$ and
$\zeta_0$, shown in Fig. \ref{fig:zeta_i}. Taking into account the
constraint on the inclination angle, $20<i<60$ \citep{casares05}, we find
that the opening angle of the hollow cone anisotropy pattern  is constrained
to  $40^\circ<\zeta_0<75^\circ$, if the emission is assumed to come from the vicinity of the
compact object ($\chi=0^\circ$ in Fig. \ref{fig:conus}).

\textit{ A constraint on the geometry of LS 5039.}
The opening angle of the hollow cone, $\zeta_0(E_\gamma)$, can be
found, once the emission process leading to the \gr\ production and
the location of the emission region are known. Comparing the
theoretically predicted value of $\zeta_0(E_\gamma)$ to the one
implied by the data one can, in principle, constrain the geometry of
the system. In particular, assuming that the location of the emission
region is known, one can constrain the inclination of the binary orbit
with respect to the line of sight. Otherwise, if the inclination of
the orbit would be known, one would be able to constrain the location
of the emission region, in particular, its distance from the star
and/or elevation above the orbital plane.

Different locations of the VHE \gr\ emission region are assumed in
different models of activity of LS 5039. In the model of "compact
pulsar wind nebula" (see e.g. \citep{dubus07})
the emission region is assumed to surround the compact object ($h=0$
in notations of Fig. \ref{fig:conus}). In a "microquasar" model the
TeV emission is assumed to be produced in a jet, so that the emission
region is displaced from the position of the compact object (it is
situated above or below the orbital plane, $h\not= 0$, see
e.g. \citep{khangulyan07}).  In both types of models the VHE \gr s are
produced via the inverse Compton scattering of the soft
photons by the VHE electrons.

To find the rate of production of \gr s via the inverse Compton
scattering, one has to numerically integrate the Eq.  (\ref{eq:R}),
with $\sigma_{\rm prod}$ being the Klein-Nishina cross section, over
the angular and energy distributions of soft photons at a distance $D$
from the massive star.  The result of such integration is shown by the
dotted curves in Fig. \ref{fig:Fzeta_num} for electron energies,
$E_e=1$ and $10$~TeV for the case when the emission region is situated
at the distance of the periastron of the binary orbit.

To find the optical depth $\tau_{\rm abs}(\zeta)$ for the \gr s
emitted at different angles $\zeta$ (see Eq. \ref{F}), one first
calculates the \gr\ absorption rate, given by Eq. (\ref{eq:R}) with
the cross section $\sigma$ being the pair production cross-section, at
each point of the \gr\ trajectory. This is done similarly to the
calculation of the inverse Compton scattering rate, via an integration
over the soft photon angular distribution. Next, one has to integrate
the absorption rate along the \gr\ trajectory, from the emission point
to infinity. The result of such numerical integration is shown by the
dashed curves in Fig. \ref{fig:Fzeta_num} for \gr\ energies
$E_\gamma=1$ and $10$~TeV, assuming that emission is produced at the
distance of the periastron of the binary orbit.

\begin{figure}
\plotone{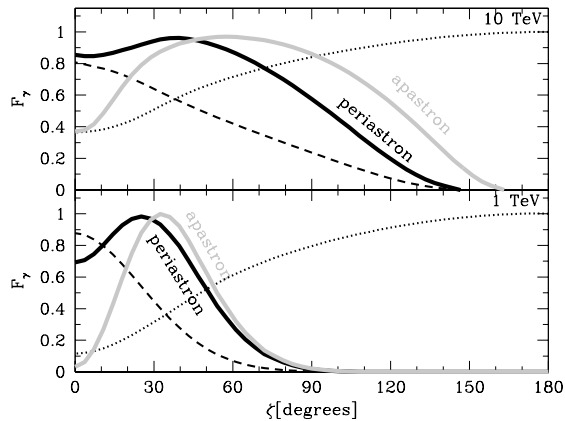}
  \caption{Angular brightness profile of the hollow cone
  in LS 5039, found assuming that emission is produced at the distance of 
  periastron (black) or apastron (grey) of the orbit. Thin dotted and dashed
  curves show the production rate and absorption coefficient for an emission
  region at the periastron distance.}
\label{fig:Fzeta_num}
\end{figure}

Substituting the numerically found ${\cal R}_{\rm
  prod}(\zeta,E_\gamma)$ and $\tau_{\rm abs}(\zeta, E_\gamma)$ into
Eq. (\ref{F}) one can find the angular brightness profile of the
hollow cone for different orbital phases and different photon
energies. The angular brightness profile, calculated for the
periastron/apastron of the binary orbit, is shown by the thick, solid,
black/grey curve in Fig. \ref{fig:Fzeta_num}. We assume that the
energies of \gr s are approximately equal to the energies of the
primary electrons, which is true in the Klein-Nishina regime of the
inverse Compton scattering.

The first maximum of the TeV band lightcurve of LS 5039 takes place at the
orbital phase $\Phi_1\simeq 224^\circ-35^\circ\simeq 193^\circ$, close to the
apastron of the orbit. From Fig. \ref{fig:Fzeta_num} one can see that at this
phase the opening angle of the cone changes  in the range 
$30^\circ<\zeta_0(E_\gamma)<60^\circ$ when the \gr\ energy changes from 1 to 10
TeV.  Comparing this numerically calculated range of values (shaded region in
Fig. \ref{fig:zeta_i}) to the one implied by the observational data, one can
find from Fig. \ref{fig:zeta_i}, that if the VHE \gr\ emission is produced close
to the compact object (i.e. at $\chi=0^\circ$) the inclination angle of the 
binary orbit
should be  $i>40^\circ$.

The constraint on $i,\chi$ can be re-formulated in a different
way: if the inclination angle of the orbit is small,
e.g. $i\sim 25^\circ$ (see \citet{casares05}), the
 \gr\ emission is not produced close to the compact
object. Instead, from Fig. \ref{fig:zeta_i} one can find that in this
case the elevation $\chi$ of the emission region above the orbital
plane should be $\chi\ge 15^\circ$. If the emission region is located
in a jet-like outflow orthogonal to the orbital plane, the emission
region should be situated at the height $h=D\mbox{ tan}\chi>0.3D$
above the orbital plane, where $D$ is the binary separation distance
(see Fig. \ref{fig:conus} for notations).

\textit{ Summary.} We have shown that VHE \gr\ emission from \gr -loud
binaries is expected to have a rotating hollow cone anisotropy
pattern, determined by the Doppler effect in the anisotropic radiation
field of a massive star (Fig. \ref{fig:Fzeta}). This anisotropy leads
to the appearance of a double-peak structure of the orbit-folded
lightcurve, with the two peaks situated at equal distance $\Delta\Phi$
(Eq. \ref{DP}) from the phase of the inferior conjunction. We have
demonstrated that such a symmetric double-peak structure is observed
in the particular case of LS 5039. In this case, a measurement
of the phase shift $\Delta \Phi$ enables to find a relation between
the opening angle of the hollow cone, $\zeta_0$, and the inclination
of the binary orbit, $i$ (see Fig. \ref{fig:zeta_i}). Comparing the
value of $\zeta_0$, inferred from the data, to the one found from
numerical calculation of the angular brightness profile of the cone
(Fig. \ref{fig:Fzeta_num}) we were able to constrain the inclination
of the binary orbit and/or the elevation of the VHE \gr\ emission
region above the orbital plane in this particular source.

In the particular case of LS 5039, the rotating hollow cone model
discussed above can be tested if the statistics of the signal from the
source becomes high enough to allow a splitting of the source
lightcurve at the energies $E>1$~TeV onto two energy bins
(e.g. $1$~TeV$<E<10$~TeV and $E>10$~TeV). In this case the predicted
shifts of the two maxima of the lightcurve toward each other (or even
a merger of them) at the higher energies should be observable. If
observed, such an effect would be a clear evidence in favor of the
proposed model.  

We thank F.Aharonian and A.Zdziarsky for the
discussions of the subject.

\end{document}